\documentstyle[11pt,twoside]{article}

\evensidemargin=\oddsidemargin
\makeatletter
\def\LegE{\begin{picture}(0,0)
   \put( 0.25,    0){\vector( 1, 0){0.50}}
   \@ifstar{\@flE}{\@@flE}}
\def\@flE  #1{\put( 0.5 ,-0.03){\makebox(0,0)[ t]{$#1$}}\end{picture}}
\def\@@flE #1{\put( 0.5 , 0.03){\makebox(0,0)[ b]{$#1$}}\end{picture}}
\def\LegNE{\begin{picture}(0,0)
   \put( 0.18, 0.18){\vector( 1, 1){0.64}}
   \@ifstar{\@flNE}{\@@flNE}}
\def\@flNE #1{\put( 0.52, 0.48){\makebox(0,0)[tl]{$#1$}}\end{picture}}
\def\@@flNE#1{\put( 0.48, 0.52){\makebox(0,0)[br]{$#1$}}\end{picture}}
\def\LegN{\begin{picture}(0,0)
   \put(    0, 0.20){\vector( 0, 1){0.60}}
   \@ifstar{\@flN}{\@@flN}}
\def\@flN  #1{\put( 0.03, 0.5 ){\makebox(0,0)[ l]{$#1$}}\end{picture}}
\def\@@flN #1{\put(-0.03, 0.5 ){\makebox(0,0)[ r]{$#1$}}\end{picture}}
\def\LegNW{\begin{picture}(0,0)
   \put(-0.18, 0.18){\vector(-1, 1){0.64}}
   \@ifstar{\@flNW}{\@@flNW}}
\def\@flNW #1{\put(-0.48, 0.52){\makebox(0,0)[bl]{$#1$}}\end{picture}}
\def\@@flNW#1{\put(-0.52, 0.48){\makebox(0,0)[tr]{$#1$}}\end{picture}}
\def\LegW{\begin{picture}(0,0)
   \put(-0.25,    0){\vector(-1, 0){0.50}}
   \@ifstar{\@flW}{\@@flW}}
\def\@flW  #1{\put(-0.5 , 0.03){\makebox(0,0)[ b]{$#1$}}\end{picture}}
\def\@@flW #1{\put(-0.5 ,-0.03){\makebox(0,0)[ t]{$#1$}}\end{picture}}
\def\LegSW{\begin{picture}(0,0)
   \put(-0.18,-0.18){\vector(-1,-1){0.64}}
   \@ifstar{\@flSW}{\@@flSW}}
\def\@flSW #1{\put(-0.52,-0.48){\makebox(0,0)[br]{$#1$}}\end{picture}}
\def\@@flSW#1{\put(-0.48,-0.52){\makebox(0,0)[tl]{$#1$}}\end{picture}}
\def\LegS{\begin{picture}(0,0)
   \put(    0,-0.2 ){\vector( 0,-1){0.60}}
   \@ifstar{\@flS}{\@@flS}}
\def\@flS  #1{\put(-0.03,-0.5 ){\makebox(0,0)[ r]{$#1$}}\end{picture}}
\def\@@flS #1{\put( 0.03,-0.5 ){\makebox(0,0)[ l]{$#1$}}\end{picture}}
\def\LegSE{\begin{picture}(0,0)
   \put( 0.18,-0.18){\vector( 1,-1){0.64}}
   \@ifstar{\@flSE}{\@@flSE}}
\def\@flSE #1{\put( 0.48,-0.52){\makebox(0,0)[tr]{$#1$}}\end{picture}}
\def\@@flSE#1{\put( 0.52,-0.48){\makebox(0,0)[bl]{$#1$}}\end{picture}}
\def\capsa(#1,#2)#3{\put(#1,#2){\makebox(0,0){$#3$}}}
\def\diagr{\@ifnextchar [{\@diagr}{\@diagr[15ex]}}
\def\@diagr[#1](#2,#3){\begingroup
   \setlength{\unitlength}{#1}
   \begin{picture}(#2,#3)}
\def\enddiagr{\end{picture}
   \endgroup}
\def\indiag{\@ifnextchar [{\@indiag}{\@indiag[15ex]}}
\def\@indiag[#1](#2,#3){\begingroup
   \setlength{\unitlength}{#1}
   \medskip
   \begin{center}
   \begin{picture}(#2,#3)}
\def\exdiag{\end{picture}
   \end{center}
   \medskip
   \endgroup}
\def\fflE{\begin{picture}(0,0)
   \put( 0.25,    0){\vector( 1, 0){1.50}}
   \@ifstar{\@fflE}{\@@fflE}}
\def\@fflE  #1{\put( 1   ,-0.03){\makebox(0,0)[ t]{$#1$}}\end{picture}}
\def\@@fflE #1{\put( 1   , 0.03){\makebox(0,0)[ b]{$#1$}}\end{picture}}
\def\qed{\ifvmode\removelastskip\fi
{\unskip\nobreak\hfil\penalty50\hbox{}\nobreak\hfil
\hbox{\vrule height1.2ex width1.2ex}\parfillskip=0pt
\finalhyphendemerits=0 \par\smallskip}}

\def\dif{{\rm d}}
\def\deriv{\@ifnextchar[{\@deriv}{\@deriv[]}}
   \def\@deriv[#1]#2#3{\mathchoice%
{{\dif^{#1}#2\over\dif{#3}^{#1}}}{{\dif^{#1}#2/\dif{#3}^{#1}}}%
{{\dif^{#1}#2\over\dif{#3}^{#1}}}{{\dif^{#1}#2/\dif{#3}^{#1}}}}
\def\derpar#1#2{\mathchoice%
{{\partial#1\over\partial#2}}{{\partial#1/\partial#2}}%
{{\partial#1\over\partial#2}}{{\partial#1/\partial#2}}}
\def\dderpar#1#2#3{\mathchoice%
{{\partial^2 #1\over\partial #2\,\partial #3}}%
{{\partial^2 #1/\partial #2\,\partial #3}}%
{{\partial^2 #1\over\partial #2\,\partial #3}}%
{{\partial^2 #1/\partial #2\,\partial #3}}}

\def\secteqno{\@addtoreset{equation}{section}%
\def\theequation{\thesection.\arabic{equation}}}
\newcounter{subequation}
\def\thesubequation{\alph{subequation}}
\def\sneqnarray{\stepcounter{equation}\let\@currentlabel=\theequation
\setcounter{subequation}{1}
\def\@eqnnum{{\rm (\theequation.\thesubequation)}}
\global\@eqcnt\z@\tabskip\@centering\let\\=\@eqncr\let\@@eqncr=\@@sneqncr
$$\halign to \displaywidth\bgroup\@eqnsel\hskip\@centering
 $\displaystyle\tabskip\z@{##}$&\global\@eqcnt\@ne
 \hskip 2\arraycolsep \hfil${##}$\hfil
 &\global\@eqcnt\tw@ \hskip 2\arraycolsep $\displaystyle\tabskip\z@{##}$\hfil
  \tabskip\@centering&\llap{##}\tabskip\z@\cr}
\def\endsneqnarray{\@@sneqncr\egroup $$\global\@ignoretrue}
\def\@@sneqncr{\let\@tempa\relax
   \ifcase\@eqcnt \def\@tempa{& & &}\or \def\@tempa{& &}
   \else \def\@tempa{&}\fi
     \@tempa \if@eqnsw\@eqnnum\stepcounter{subequation}\fi
     \global\@eqnswtrue\global\@eqcnt\z@\cr}
\makeatother
\def\artit#1{``#1'',}
\def\ben{\begin{enumerate}}
\def\een{\end{enumerate}}
\def\beq{\begin{equation}}
\def\eeq{\end{equation}}
\def\bea{\begin{eqnarray}}
\def\eea{\end{eqnarray}}
\def\beann{\begin{eqnarray*}}
\def\eeann{\end{eqnarray*}}
\def\beasn{\begin{sneqnarray}}
\def\eeasn{\end{sneqnarray}}

\newtheorem{teor}{Theorem}
\newtheorem{prop}{Proposition}

\newtheorem{cor}{Corollary}

\let\ds=\displaystyle
\def\buildord#1\over#2{\mathord{\mathop{\kern0pt #2}\limits^{#1}}}

\def\map#1{\mathrel{\mathop{\to}\limits^{#1}}}
\def\mapping#1{\mathrel{\mathop{\longrightarrow}\limits^{#1}}}
\def\inmap#1{\mathrel{\mathop{\hookrightarrow}\limits^{#1}}}

\def\Id{{\rm Id}}

\def\Real{{\bf R}}
\let\isom=\cong

\def\Ker{\mathop{\rm Ker}\nolimits}
\def\Hom{\mathop{\rm Hom}\nolimits}
\def\Lin{{\cal L}}
\def\transp#1{{}^{t}\kern-.15em\relax#1}

\def\lop{\!\cdot\!}

\def\Dif{{\rm D}}

\def\feble#1{\mathrel{\mathop{\approx}\limits_{#1}}}
\def\forta#1{\mathrel{\mathop{\cong}\limits_{#1}}}

\def\comp{\mathbin{\scriptstyle\circ}}
\def\scomp{\mathbin{\scriptstyle\bullet}}
\def\opers#1{(\!\!(#1)\!\!)}

\def\Tan{{\rm T}}
\def\Ver{{\rm V}}

\def\vl{{\rm vl}}
\def\Lio{\Delta}
\def\vend{{\rm J}}

\def\calF{{\cal F}}

\def\sId{\mathord{\cal I\!\it d}}
\def\Der{{\cal F}}
\def\vv{^{\rm v}}

\def\ELform{{\cal E}_L}
\def\Leg{\Der \!L}
\def\Lleg{\Der^2\!L}
\def\En{E_L}
\def\gam{\gamma}
\def\Gam{{\mit\Gamma}}
\def\Del{{\mit\Delta}}
\def\Ups{{\mit\Upsilon}}
\def\lam{v}

\def\XL{{X}^{\!^{\rm L}}}
\def\XH{{X}^{\!^{\rm H}}}

\let\eps\varepsilon
\def\proof{\noindent{\it Proof}.\quad}

\title{\sf Singular lagrangians: some geometric structures\\ 
along the Legendre map%
\thanks{published: 
{\sl J.~Phys.~A: Math.\ Gen.~\bf 34} (2001) 3047--3070}
}

\def\tabaddress#1{{\small\it\begin{tabular}[t]{c}#1\\[1.2ex]\end{tabular}}}
\def\UPCMAT{Departament de Matem\`atica Aplicada IV\\
   Universitat Polit\`ecnica de Catalunya\\
   Campus Nord UPC, edifici C3\\
   C. Jordi Girona, 1\\
   08034 Barcelona\\
   Catalonia, Spain}
\def\UBECM{Departament d'Estructura i Constituents de la Mat\`eria,
   Universitat de Barcelona\\
   and Institut de F\'{\i}sica d'Altes Energies\\
   Av.~Diagonal 647\\
   08028 Barcelona\\
   Catalonia, Spain}

\author{\sf 
Xavier Gr\`acia$^a$ and Josep M. Pons$^b$
\\[2mm]
\tabaddress{$^a$\UPCMAT}
\\
\tabaddress{$^b$\UBECM}
\\[2mm]
\small\sf emails: 
xgracia@mat.upc.es, 
pons@ecm.ub.es
}

\date{\sf 29 September 2000 / updated April 2001}

\secteqno

\begin{document}
\abovedisplayskip=6pt plus 3pt minus 1pt
\belowdisplayskip=\abovedisplayskip
\belowdisplayshortskip=4pt plus 3pt minus 1pt

\maketitle
\thispagestyle{empty}
\vskip 4mm

\begin{abstract}
\noindent
New geometric structures that relate
the lagrangian and hamiltonian formalisms 
defined upon a singular lagrangian
are presented.
Several vector fields are constructed in velocity space
that give new and precise answers to several topics like
the projectability of a vector field to a hamiltonian vector field,
the computation of the kernel of 
the presymplectic form of lagrangian formalism,
the construction of the lagrangian dynamical vector fields,
and the characterisation of dynamical symmetries.
\bigskip
\parindent 0pt
\it


\smallskip

MSC\,2000: 70H45, 70G45 
\qquad
PACS\,1999: 45.20.Jj, 02.40.Vh

\end{abstract}
\newpage

\section{Introduction}

The dynamics associated with
a first-order time-independent variational principle 
on a configuration manifold~$Q$
can be formulated either
in its tangent bundle $\Tan Q$ (lagrangian formalism)
or in its cotangent bundle $\Tan^*Q$ (hamiltonian formalism).
If the variational problem is defined by the lagrangian function~$L$,
both formulations are related through the Legendre transformation,
which is given by the fibre derivative of~$L$,
$\Leg \colon \Tan Q \to \Tan^*Q$.

In the regular case, that is, 
when $\Leg$ is a local diffeomorphism
(or when the fibre hessian is everywhere non-singular),
the equivalence between both formulations is fairly simple.
However, in the singular case, 
this correspondence between the lagrangian and the hamiltonian formalisms 
is far from trivial,
and it is just this case which is 
the most relevant for the fundamental physical theories
(as generally covariant theories, Yang-Mills theories and string theory),
because the occurrence of gauge freedom 
is only possible within this framework. 
This explains the effort made since 1950
to define the lagrangian and hamiltonian formalisms
in the singular case,
to study the relations between them,
their dynamics and symmetries, their quantisation, and so on.
In contrast to the regular case,
some specific features of the singular case include
constraints, arbitrary functions, gauge invariance, gauge fixing, etc. 

This development has benefitted from the introduction of
differential-geometric methods in the study of dynamical systems
---some books along this line are for instance
\cite{AM-mech}
\cite{Arn-mech}
\cite{God-mec}
\cite{JS-dyn}.
A great variety of tools from differential geometry
---manifolds and bundles, differential forms, metrics, connections \ldots---
has been widely applied since the 70s to singular lagrangians,
achieving a fair comprehension about the lagrangian and the hamiltonian
formalisms and their relations.

The need of fine tools in the singular case is 
a direct consequence of the Legendre transformation 
$\Leg \colon \Tan Q \to \Tan^*Q$
being singular.
For instance, if $\Leg$ is a diffeomorphism, 
a hamiltonian vector field $Z$ in $\Tan^*Q$
(with respect to the canonical symplectic form $\omega_Q$)
is directly converted into
a hamiltonian vector field $Y = \Leg^*(Z)$ in $\Tan Q$
(with respect to the symplectic form $\omega_L = \Leg^*(\omega_Q)$,
which indeed can be used to describe the lagrangian dynamics).
In the singular case, each part of this statement
(which of course is not true)
has to be scrutinised carefully.

The purpose of this paper is to introduce 
some as yet unveiled geometric structures 
that appear in these formalisms and that facilitate the 
connection between the lagrangian and the hamiltonian formulations
in the singular case.
Once the lagrangian function is fixed,
a vector field $Y_h$ in $\Tan Q$
will be defined from an arbitrary function $h$ in $\Tan^*Q$;
this is our main object.
From it, 
once a hamiltonian and a basis for the primary hamiltonian constraints
are chosen,
another vector field $\Del_h$ will be defined;
should the lagrangian be regular,
the vector field $\Del_h$ 
would be the hamiltonian vector field of $\Leg^*(h)$
with respect to~$\omega_L$.
These constructions,
and other ones related to them,
provide new connections between the dynamics in both pictures.
Applications include the study of 
the projectability of a vector field in lagrangian formalism
to a hamiltonian vector field,
the construction of the lagrangian dynamical vector fields,
the study of the relation between 
the arbitrary functions of the lagrangian and hamiltonian dynamics, 
and the formulation of the dynamical symmetries
(with special emphasis on the Noether symmetries); 
even the intrinsic construction of some structures 
as the kernel of the presymplectic form in tangent space
will become almost trivial.

As for the geometric tools used in the paper,
they are related with the fibred structure 
of the tangent and cotangent bundles.
We use basically the fibre derivative
(that is, the ordinary differentiation with respect to
the fibre variables),
the vertical lift
(that is, 
the identification between points and tangent vectors in a vector space),
and the canonical structures of the tangent bundle
(vertical endomorphism, canonical involution)
and of the cotangent bundle
(the canonical differential forms).

The paper is organised as follows.
Sections 2 and~3 provide some differential-geometric preliminaries
concerning bundles and the fibre derivative.
Section~4 contains a geometric description 
of lagrangian and hamiltonian formalisms in the singular case.
The construction of the vector field $Y_h$ is presented in section~5,
together with some of its properties.
Two other vector fields, $R_h$ and $\Del_h$,
are also presented there.
Section~6 uses the mentioned constructions 
to study the projectability to hamiltonian vector fields
of $\Tan^*Q$,
and to give an explicit basis for the kernel of the presymplectic form
$\omega_L$ of lagrangian formalism.
In section~7 the preceding vector fields 
are used to construct the lagrangian dynamics
and to relate the arbitrary functions of lagrangian and hamiltonian
dynamics;
the dynamical symmetries of hamiltonian formalism
are also studied in a simple way.
The case of regular lagrangians is studied in section~8.
Section~9 contains a simple example.
The final section is devoted to conclusions.

\section{Some facts about bundles}

Basic techniques concerning fibre bundles and vector bundles
will be needed;
in particular, the vertical vectors of a bundle and
the tangent bundle of a bundle,
as well as some canonical structures related to the tangent bundle.
They may be found in many books, such as for instance
\cite{AM-mech}
\cite{AMR-manif}
\cite{Die-ea3}
\cite{God-mec}
\cite{KMS-natural}
\cite{Sau-jets}.
In this section we recall a few of these concepts
and introduce some notation.

\subsubsection*{Vertical vectors}

Let $\pi \colon E \to B$ be a fibre bundle,
with fibres $E_x = \pi^{-1}(x)$.
The {\it vertical bundle}\/ of $E$ is the vector subbundle
$\Ver(E) = \Ker \Tan(\pi) \subset \Tan(E)$.
Its fibre at a point $e_x \in E_x$ is the tangent space to the fibre
of~$E$ at~$x$:
$\Ver_{e_x}(E) = \Tan_{e_x}(E_x)$.

\medskip

Let us consider a {\it vector bundle}\/ $E \to B$.
At each $x \in B$ we have a vector space~$E_x$.
The tangent space of $E_x$ at a point $e_x$ is
naturally isomorphic to $E_x$ itself,
$E_x \map{\isom} \Tan_{e_x}(E_x)$;
this isomorphism is constructed by sending 
$v_x$ to the tangent vector of the path
$t \mapsto e_x+tv_x$ in~$E_x$.
Therefore
$\Tan(E_x) \isom E_x \times E_x$.

Globally this yields a canonical isomorphism
$\Ver(E) \isom E \times_B E$,
called the {\it vertical lift}
\vskip -8mm
\bea
E \times_B E & \mapping{\vl_E} & \Ver(E) \subset \Tan(E) 
\\
(e_x,v_x)    & \mapsto         & \vl_E(e_x,v_x) = [t \mapsto e_x+tv_x] 
\nonumber
\eea
Here $E \times_B E$ denotes the fibre product
(its elements are the couples $(e,e') \in E \times E$ such that
$\pi(e) = \pi(e')$),
considered as a vector bundle over the first factor.

\smallskip

The vertical lift defines a natural bijection between
fibre bundle maps $E \to E$ and vertical vector fields on~$E$:
if $\xi \colon E \to E$ is a fibre bundle map,
then the map
\beq
\xi\vv \colon
E \mapping{} \Ver(E) \subset \Tan(E) ,
\qquad
\xi\vv(e) = \vl_E (e,\xi(e)) 
\label{camp-vert}
\eeq
is a vertical vector field.
This procedure applied to the identity map of~$E$
yields a canonical vertical vector field,
the {\it Liouville's vector field}, $\Lio_E(e) = \vl_E(e,e)$.
If $(x,a)$ are vector bundle coordinates of~$E$
---usually we will omit indices---
then the local expression of $\Lio_E$ is $a^i \derpar{}{a^i}$.

\subsubsection*{Some structures of $\Tan(\Tan B)$}

Given a vector bundle $\pi \colon E \to B$,
the tangent bundle $\Tan E$ has two vector bundle structures:
$\tau_E \colon \Tan E \to E$ and 
$\Tan \pi \colon \Tan E \to \Tan B$.
In the case of $E = \Tan B$,
we obtain two different vector bundle structures over the same base.
Both structures are canonically isomorphic through the 
{\it canonical involution},
$\kappa_B \colon \Tan(\Tan B) \to \Tan(\Tan B)$.
Its local expression in natural coordinates is
\vskip -8mm
$$
\kappa(x,v;u,a) = (x,u;v,a) .
$$

Another map in this manifold is the {\it vertical endomorphism}\/
$\vend \colon  \Tan(\Tan B) \to \Tan(\Tan B)$,
whose local expression is
\vskip -8mm
$$
\vend(x,v;u,a) = (x,v;0,u) .
$$

\subsubsection*{Projectability}

Let $\calF \colon M \to N$ be a map between manifolds.
A function $f \colon M \to \Real$ is said to be 
{\it projectable}\/ (through~$\calF$)
if $f = \calF^*g := g \comp \calF$
for a certain function $g \colon N \to \Real$.
A vector field $X$ on $M$ is {\it projectable}\/ 
if there exists a vector field $Y$ on $N$ such that
$\Tan(\calF) \comp X = Y \comp \calF$;
one also says that $X$ and $Y$ are $\calF$-related.
Alternatively,
one has $X \lop \calF^*(g) = \calF^*(Y \lop g)$
for any function $g$ on~$N$.

When $\calF$ has {\it constant rank},
one can use the rank theorem to obtain a characterisation of 
the {\it local}\/ projectability of a function~$f$:
this condition is that $v \lop f = 0$
for every $v \in \Ker \Tan(\calF)$.
There are similar results for the local projectability of vector fields.
However, let us just point out one result from the opposite side:
a vector field $Y$ on $N$ is locally the projection of a vector field $X$
iff $Y$ is tangent to the image of~$\calF$.

\section{Fibre derivatives}

The fibre derivative will play an important role in our developments.
Its definition can be found in many places such as, for instance, 
\cite{GS-variations}
\cite{AM-mech},
since it is a relevant structure when constructing the Legendre
transformation that connects lagrangian and hamiltonian formalisms.
In a recent article
\cite{Gra-fibder}
the fibre derivative has been studied in detail,
with a view to application to singular lagrangian dynamics.
In this section we summarise some of the results of this paper.

\subsubsection*{Definition of the fibre derivative}

Our framework consists of 
two real {\it vector bundles}\/ $E \to M$ and $F \to M$
over the same base,
and a {\it fibre $M$-bundle morphism}\/
$f \colon E \to F$,
that is, a fibre-preserving map:
for each $e_x \in E_x$,
$f(e_x) \in F_x$.
(In
\cite{Gra-fibder}
the more general case of $E$ and $F$ being affine bundles is considered;
this is especially interesting, for instance, 
when considering higher-order or time-dependent lagrangians,
or field theory.)

The restriction of $f$ to a fibre defines a map
$f_x \colon E_x \to F_x$
between vector spaces,
whose ordinary derivative 
at a point $e_x \in E_x$ is a linear map
$\Dif f_x(e_x) \colon E_x \to F_x$.
In other words,
we have defined an element
\beq
\Der \!f(e_x) := \Dif f_x(e_x) \in \Hom(E_x,F_x)
\eeq
for each $e_x \in E$.
Globally, this defines a fibre-preserving map
\beq
\Der \!f \colon E \mapping{} \Hom(E,F) \isom F \otimes E^* ,
\eeq
which is the {\it fibre derivative}\/ of~$f$.

If the local expression of $f$ is
$(x^\mu,a^i) \mapsto (x^\mu,f^k(x,a))$,
then the local expression of $\Der \!f$ is
\beq
\Der \!f(x^\mu,a^i) = \left( x^\mu, \derpar{f^k}{a^i}(x,a) \right) .
\eeq

Since $\Der \!f$ is also a fibre bundle map between vector bundles,
the same procedure can be applied to compute its fibre derivative.
The canonical isomorphism
$\Hom(E,\Hom(E,F)) \isom \Lin^2(E;F)$
now yields the second fibre derivative,
the {\it fibre hessian},
which is the map
\beq
\Der^2 \!f  \colon E \mapping{} \Lin^2(E;F)
\isom \Hom(E \otimes E,F)
\isom F \otimes E^* \otimes E^* ,
\eeq
whose local expression is
\beq
\Der^2 \!f (x^\mu,a^i) = \left( x^\mu, \dderpar{f^k}{a^i}{a^j}(x,a) \right) .
\eeq
This can be readily generalised to higher order fibre derivatives.

\subsubsection*{The case of a real function}

Let us notice the particular case where $F = M \times \Real$.
This corresponds indeed to considering a real function
$f \colon E \to \Real$
on a vector bundle $\pi \colon E \to M$.
Then its fibre derivative is a map
\beq
\Der \!f \colon E \mapping{} \Hom(E,M \times \Real) =: E^* ,
\eeq
of which we shall study some properties.

First, there is a close relation between the tangent map
$$
\Tan(\Der \!f) \colon
 \Tan E \mapping{} \Tan E^*
$$
and the fibre hessian $\Der^2 \!f $ of~$f$,
$$
\Der^2 \!f = \Der (\Der \!f) \colon
 E \mapping{} \Hom(E,E^*)  \isom  E^* \otimes E^* .
$$
Indeed, the restriction of $\Tan_{e_x}(\Der \!f)$ to vertical vectors is
---thanks to the vertical lift---
essentially the same map as the hessian considered as a map
$\Der^2 \!f(e_x) \colon E_x \to E_x^*$.
As a consequence, one has that
$$
v_x \in \Ker \Der^2 \!f(e_x) 
\iff 
\vl_E(e_x,v_x) \in \Ker \Tan_{e_x}(\Der \!f) ,
$$
and since
$
\Ker \Tan(\Der \!f) \subset \Ver(E) 
$,
in this way we obtain the whole subbundle $\Ker \Tan(\Der \!f)$.
Notice in particular that
$\Der \!f$ is a local diffeomorphism at $e_x \in E$ iff
$\Der^2 \!f(e_x)$ is a linear isomorphism.

These results can be also deduced from the local expressions of the maps;
using as natural coordinates of $E$ and $E^*$
$(x,a)$ and $(x,\alpha)$ respectively, they are:
\beann
\Der \!f : &&
(x,a) \mapsto \left(x, \derpar{f}{a}(x,a) \right) ,
\\
\Tan(\Der \!f) : &&
(x,a;v,h) \mapsto 
\left(x,\derpar{f}{a}(x,a); v, \dderpar fax v + \dderpar faa h \right) ,
\\
\Der^2 \!f : &&
(x,a) \mapsto \left(x, \dderpar faa(x,a) \right) .
\eeann

\medskip

Finally we want to notice the following result.
If $\xi \colon E \to E$ is a bundle map
with associated vertical vector field $X = \xi\vv$ on~$E$,
and $g \colon E \to \Real$ is a function,
then 
\beq
X \lop g = \langle \Der g, \xi \rangle .
\label{Lie-ver}
\eeq
This can be applied in particular to
the Liouville's vector field, giving
\beq
(\Lio_E \lop g)(e_x) = \langle \Der g(e_x),e_x \rangle ;
\label{Lio}
\eeq
the fibre derivative of this expression can be computed 
by applying the Leibniz's rule, and is
\beq
\Der(\Lio_E \lop g)(e_x) = \Der g(e_x) + \Der^2g(e_x) \lop e_x .
\label{D-Lio}
\eeq

\subsubsection*{Some useful structures: $\Gam_h$ and $\Ups^g$}

Considering the fibre derivative
$\Der \!f \colon E \to E^*$ of~$f$
as fixed data,
we are going to derive several properties of a function
$h \colon E^* \to \Real$ and its fibre derivatives.

We use the notation
\beq
\gam_h = \Der h \comp \Der \!f \colon E \to E
\label{gam_h}
\eeq
for the composition
$
E \mapping{\Der \!f} E^* \mapping{\Der h} E^{**} \isom E
$.
Recall that this map,
through the vertical lift, 
defines a vertical vector field $\gam_h\vv$ on~$E$:
\beq
\Gam_h := \gam_h\vv = \vl_E \comp (\Id_E, \Der h \comp \Der \!f)
\colon E \to E \times_M E \to \Ver\! E \subset \Tan E .
\label{Gam_h}
\eeq
Their local expressions are
$$
\gam_h \colon (x,a) \mapsto \left(x,\derpar{h}{\alpha}(\Der \!f(x,a))\right)
\qquad
\Gam_h = (\Der \!f)^* \left( \derpar{h}{\alpha_i} \right) \derpar{}{a^i} .
$$

We can apply the chain rule to compute expressions like
\beq
\Der (h \comp \Der \!f) = \Der^2 \!f \scomp \gam_h ,
\label{D-h-F}
\eeq
\beq
\Der (\gam_h) = (\Der^2 h \comp \Der \!f) \scomp \Der^2 \!f .
\label{D-gamh}
\eeq
Here we have, for instance,
$
\Der^2 h \comp \Der \!f \colon
E \to E^* \to \Hom(E^*,E^{**}) \isom \Hom(E^*,E) 
$
and
$\Der^2 \!f \colon E \to \Hom(E,E^*)$;
the symbol $\scomp$ denotes 
the composition between the images of both maps
---it is like the contraction of vector fields with differential forms.

\smallskip

Notice from (\ref{D-h-F})
that if $h$ vanishes on the image $\Der \!f(E) \subset E^*$
then $\gam_h$ is in the kernel of~$\Der^2 \!f$.
So we obtain the following result
---see also
\cite{Gra-fibder}
\cite{BGPR-equiv}:

{\it
Suppose that $\Der \!f$ has constant rank;
thus, locally the image of $\Der \!f$ is a submanifold of~$E^*$
that can be (locally) described by the vanishing of
a set of independent functions 
$\phi_\mu \colon E^* \to \Real$.
Then the vectors $\gam_{\phi_\mu}(e_x)$ are a basis for 
$\Ker \Der^2 \!f(e_x)$,
and the vertical vector fields $\Gam_{\phi_\mu}$
constitute a frame for $\Ker \Tan(\Der \!f)$.
}

As a byproduct, a function on $E$ 
is (locally) projectable through $\Der \!f$ to $E^*$ iff 
its Lie derivative with respect to the vector fields $\Gam_{\phi_\mu}$
is zero.

\bigskip

Now we present a construction dual to~$\Gam_h$.
Given a function $g \colon E \to \Real$, 
we can use its fibre derivative $\Der g \colon E \to E^*$
to construct a map
\beq
\Ups^g = \vl_{E^*} \comp (\Der \!f, \Der g)
\colon
E \to E^* \times_M E^* \to \Ver E^* \subset \Tan E^* ;
\label{Ups^g}
\eeq
this is a vector field along the map~$\Der \!f$,
with local expression
$$
\Ups^g = \derpar{g}{a^i} \left( \derpar{}{\alpha_i} \comp \Der \!f \right) .
$$

Recall that a section of a bundle $\pi \colon E \to B$ along a map
$f \colon B' \to B$
is a map $\sigma \colon B' \to E$
such that $\pi \comp \sigma = f$.
In particular, 
a section $Z \colon B' \to \Tan B$ of $\Tan B$ along~$f$
is called a {\it vector field along}~$f$;
such a map derivates a function $h \colon B \to \Real$ 
giving a function $Z \lop h$ on~$B'$:
$(Z \lop h)(y) = Z(y) \lop h$.

Notice finally that, as differential operators, 
$\Gam_h$ and $\Ups^g$ are related by
\beq
\Ups^g \lop h = \Gam_h \lop g .
\label{Ups-Gam}
\eeq
This is follows from the fact that
$
\Gam_h \lop g = 
\langle \Der g, \gam_h \rangle =
\langle \Der g, \Der h \comp \Der f \rangle =
\Ups^g \lop h
$.

\section{Some structures of lagrangian and hamiltonian formalisms}

The basic concepts about singular lagrangian and hamiltonian formalisms
---Legendre map, energy, hamiltonian function, hamiltonian constraints
\ldots---
are well known and can be found in several papers,
such as for instance
\cite{BGPR-equiv}
\cite{BK-pres}
\cite{Car-theory}
\cite{GNH-pres}
\cite{MMS-constraints}
\cite{MT-ham}.
Now we will recall some of these concepts,
introducing also some recent results from
\cite{Gra-fibder}.

\subsubsection*{Connection between the lagrangian and the hamiltonian spaces}

Let us consider a first-order autonomous lagrangian
on a configuration space~$Q$,
that is to say, a map
$L \colon \Tan Q \map{} \Real$.
Its fibre derivative (Legendre transformation) and fibre hessian are maps
\vskip -2mm
$$
\Leg \colon 
\Tan Q \mapping{} 
\Tan^*Q ,
$$
$$
\Lleg = \Der (\Leg) \colon
\Tan Q \mapping{} 
\Hom(\Tan Q,\Tan^*Q) = \Tan^*Q \otimes \Tan^*Q .
$$
The local expression of $\Leg$ is
$\Leg(q,\dot q)=(q,\hat p)$,
where $\ds \hat p = \derpar{L}{\dot q}$ are the momenta.
If the Legendre map is a local diffeomorphism 
---equivalently the hessian is everywhere nonsingular---
the lagrangian $L$ is called {\it regular},
otherwise it is called {\it singular}
---this is our focus of interest.

\medskip

We assume that 
the Legendre transformation of~$L$ has connected fibres and 
is a submersion onto a closed submanifold $P_o \subset \Tan^*Q$,
the {\it primary hamiltonian constraint submanifold}
---that is to say,
$L$ is an almost regular lagrangian
in the terminology of
\cite{GN-preslag}.
This is the most basic technical requirement
to develop a hamiltonian formulation from
a singular lagrangian~$L$,
though from a local viewpoint it suffices to have $\Leg$ of constant rank.
Locally $P_o$ can be described by the vanishing of 
an independent set of functions~$\phi_\mu$,
called the 
{\it primary hamiltonian constraints}.
According to the preceding section,
the vectors $\gam_\mu = \gam_{\phi_\mu}$
constitute a basis for the kernel of~$\Lleg$,
and the vertical vector fields $\Gam_\mu = \Gam_{\phi_\mu}$
constitute a frame for $\Ker \Tan(\Leg)$.

\smallskip

The {\it energy}\/ of~$L$ is defined by
$$
\En = \Lio_{\Tan Q} \lop L - L .
$$
Due to the properties of the Liouville's vector field
(\ref{Lio}) (\ref{D-Lio}),
\beq
\En(u_q) = \langle \Leg(u_q),u_q \rangle - L(u_q) ,
\eeq
\beq
\Der \En(u_q) = \Lleg(u_q) \lop u_q .
\label{D-En}
\eeq
This shows at once that 
$\Gam_\mu \lop \En = \langle \Der \En , \gam_\mu \rangle = 0$,
that is to say, 
the energy is projectable (through~$\Leg$) to a function
$H \colon \Tan^*Q \to \Real$
called a {\it hamiltonian},
$$
\En = H \comp \Leg ,
$$
which is unique on the primary hamiltonian constraint submanifold.

\subsubsection*{A resolution of the identity}

Given an almost regular lagrangian~$L$,
the choice of a hamiltonian and set of primary hamiltonian constraints
yields a (local) resolution of the identity map of~$\Tan Q$ as follows:

{\it 
There exist functions $\lam^\mu$
(defined on an open set of $\Tan Q$)
such that, locally,
\beq
\Id_{\Tan Q} = \gam_H + \sum_{\mu} \gam_\mu \, \lam^\mu .
\label{lam}
\eeq
Moreover,
\beq
\sId_{\Hom(\Tan Q,\Tan Q)} =
M \scomp \Lleg + \sum_\mu \gam_\mu \otimes \Der \lam^\mu ,
\label{IMW}
\eeq
where
\beq
M =
(\Der^2 H \comp \Leg) + \sum_\mu (\Der^2 \phi_\mu \comp \Leg) \, \lam^\mu .
\label{M}
\eeq
}
(Notice that $\Lleg$ is a map $\Tan Q \to \Hom(\Tan Q,\Tan^*Q)$
and $M$ is a map 
$\Tan Q \to \Hom(\Tan^*Q,\Tan Q) = \Tan Q \otimes \Tan Q$.)

Since the functions $\lam^\mu$ and their properties
will be instrumental throughout the paper,
we will recall the proof of this result 
\cite{Gra-fibder}.
Application of the chain rule (\ref{D-h-F}) to the definition of~$H$ yields
$
\Der \En(u_q) = \Lleg(u_q) \lop \gam_H(u_q) ,
$
and so using (\ref{D-En}) we obtain
$$
\Lleg(u_q) \lop (u_q - \gam_H(u_q)) = 0 .
$$
The terms in parentheses are in $\Ker \Lleg(u_q)$,
thus there exist numbers $\lam^\mu(u_q)$ such that
$
u_q - \gam_H(u_q) = \sum_{\mu} \gam_\mu(u_q) \,\lam^\mu(u_q)
$,
which is equation (\ref{lam}).
Finally, using (\ref{D-gamh}) and the Leibniz's rule,
one can compute the fibre derivative of (\ref{lam});
the result is equation (\ref{IMW}).

\smallskip

The above results can be given a slightly different form,
using the identification of bundle maps $\Tan Q \to \Tan Q$
with vertical vector fields on $\Tan Q$.
For instance,
equation (\ref{lam})
can be rewritten as
\beq
\Lio_{\Tan Q} = \Gam_H + \sum_{\mu} \lam^\mu \, \Gam_\mu .
\label{lam'}
\eeq

\smallskip

Notice that
application of (\ref{IMW}) to $\gam_\nu$ yields
$\gam_\nu = 
\sum_\mu \gam_\mu \langle \Der \lam^\mu, \gam_\nu \rangle$.
So we have
\beq
\Gam_\nu \lop \lam^\mu 
= \langle \Der \lam^\mu, \gam_\nu \rangle 
= \delta^{\mu}_{\,\nu} ,
\label{lam-gam}
\eeq
where we have applied equation (\ref{Lie-ver}).
This shows that the functions $\lam^\mu$ are not projectable;
in a certain sense, they correspond
to the velocities that can not be retrieved from the momenta 
through the Legendre map.

\smallskip

Let us finally remark that the local expressions of 
equations (\ref{IMW}) and (\ref{M})
were initially deduced in
\cite{BGPR-equiv}
by derivating the local expression of
(\ref{lam}), which is
$$
\dot q^i =
\Leg^* \left( \derpar{H}{p_i} \right) +
\sum_\mu \Leg^* \left( \derpar{\phi_\mu}{p_i} \right) \lam^\mu .
$$

\subsubsection*{The Euler-Lagrange equation}

So far we have not considered the equations of motion.
We will deal with them in several forms.

Let $\omega_Q$ be the canonical 2-form of $\Tan^*Q$
(in coordinates $\dif q^i \wedge \dif p_i$).
One defines the presymplectic form in $\Tan Q$
$$
\omega_L = \Leg^*(\omega_Q) 
$$
---it is a symplectic form iff the lagrangian is regular.
Then a path $\gamma \colon I \to Q$
is a solution of the Euler-Lagrange equation iff
\beq
i_{\ddot \gamma} \omega_L = \dif E_L \comp \dot \gamma .
\label{EL-pres}
\eeq

A second representation of the equation of motion is
\beq
\ELform \comp \ddot\gamma = 0 ,
\label{EL-ELform}
\eeq
where $\ELform \colon \Tan^2Q \to \Tan^*Q$ 
is the {\it Euler-Lagrange form}\/ of~$L$
---see for instance
\cite{CLM-higherNoether}
\cite{Tul75};
$\Tan^2 Q$ denotes the second-order tangent bundle of~$Q$.
$\ELform$~is a 1-form along the projection $\Tan^2Q \to Q$,
with local expression
\beq
\ELform = [L]_i \,\dif q^i ,
\quad
[L]_i = \derpar{L}{q^i} - \deriv{}{t} \left( \derpar{L}{\dot q^i} \right) .
\eeq

A third version of the Euler-Lagrange equation can be written
using the time-evolution operator $K$
that connects lagrangian and hamiltonian formalisms.
This operator was expressed in
\cite{GP-K}
as a vector field along~$\Leg$
satisfying certain properties that determine it completely.
The local expression of $K$ is
$$
K(q,\dot q) = \left( q,\widehat p; \dot q, \derpar Lq \right) .
$$
In coordinates, $K$ was first introduced 
\cite{BGPR-equiv}
as a differential operator
---see also
\cite{CL-K}
\cite{Car-theory}.
Then its local expression reads
\beq  
K \lop h = 
\Leg^*\left(\derpar hq\right) \dot q +
\Leg^*\left(\derpar hp\right) \derpar Lq .
\label{Kc'}
\eeq
(In a time-dependent framework it would hold an additional piece, 
$\Leg^*(\derpar ht)$.)
The operator $K$ is a useful tool in the theory of singular lagrangians:
it can be used ---see below--- to express the equations of motion
\cite{GP-K},
to relate the lagrangian and the hamiltonian constraints
\cite{BGPR-equiv}
\cite{CL-K}
\cite{Pon-newrel},
to study the symmetries of the equations of motion
\cite{GP-hamdst}
\cite{BGGP-noether}
\cite{FP-noether}
\cite{GP-lagdst}
\cite{GP-nocons}
\cite{GP-commute}
and, more recently, to study lagrangian systems with generic singularities 
\cite{PV-sing}.
See also
\cite{GPR-higher}
\cite{GP-higherNoether}.

Using this operator, 
a path $\xi \colon I \to \Tan Q$ is the lift $\dot\gamma$
of a solution of the Euler-Lagrange equation iff
\beq
\Tan(\Leg) \comp \dot\xi = K \comp \xi .
\label{EL-K}
\eeq
The following diagram shows all the objects involved:
\indiag(2,1.05)
\capsa(0,0){I}
\capsa(1,0){\Tan Q}
\capsa(2.1,0){\Tan^*Q}
\capsa(0.95,1){\Tan(\Tan Q)}
\capsa(2.15,1){\Tan(\Tan^*Q)}
\put(0,0){\LegE*{\xi}}
\put(1,0){\LegE*{\Leg}}
\put(0,0){\LegNE{\dot\xi}}
\put(1,0){\LegNE{K}}
\put(1,1){\LegE{\Tan(\Leg)}}
\put(1,1){\LegS{}}
\put(2.1,1){\LegS{}}
\exdiag
\noindent

\subsubsection*{The Hamilton-Dirac equation}

In the singular case, hamiltonian dynamics
was first studied by Dirac and Bergmann
\cite{Dir50}
\cite{AB-lligams}
\cite{Dir-lectures}.
A path $\eta \colon I \to P_o$ is a solution of the Hamilton-Dirac equation
if there exist functions $\lambda^\mu$ such that
\beq
\dot\eta = Z_H \comp \eta + \sum_\mu \lambda^\mu \,Z_\mu \comp \eta .
\label{HD}
\eeq 
Here we denote by $Z_h$ the hamiltonian vector field defined by~$h$:
it satisfies 
$$
i_{Z_h} \omega_Q = \dif h ,
$$
and, as a differential operator, it is related to the Poisson's bracket by
$$
Z_h = \{ -,h\} .
$$
We have also put $Z_\mu = Z_{\phi_\mu}$.

Another geometric version of Dirac's theory can be obtained by considering
$j \colon P_o \inmap{} \Tan^*Q$
and the presymplectic form $\omega_o = j^*(\omega_Q)$.
Then the Hamilton-Dirac equation for a path $\eta \colon I \to P_o$ is
\beq
i_{\dot \eta} \omega_o = \dif H_o \comp \eta ,
\label{HD-pres}
\eeq
where $H_o$ is the hamiltonian on~$P_o$
\cite{GNH-pres}
\cite{BK-pres}.

Using the operator~$K$, the Hamilton-Dirac equation can be written also as
\beq
\dot\eta = K \comp \Tan(\tau_Q^*) \comp \dot\eta 
\label{HD-K}
\eeq
for a path $\eta$ in $\Tan^*Q$
\cite{GP-K}
---see also
\cite{BGPR-equiv}
\cite{Tul76}.

\medskip

Of course, the hamiltonian dynamics is defined so as to be equivalent
to the lagrangian dynamics, in the sense that
if $\xi \colon I \to \Tan Q$ is a solution of the Euler-Lagrange equation
then $\eta \colon I \to \Tan^*Q$ 
defined as $\eta = \Leg \comp \xi$ 
satisfies the Hamilton-Dirac equation,
and conversely 
taking $\eta$ and defining 
$\xi = (\tau_Q^* \comp \eta)^{\textstyle.}$ from it.
We will say that such $\xi$, $\eta$
are a couple of related solutions.

\subsubsection*{Some further relations involving the operator~$K$}

Since the same dynamics is written in different ways,
there are relations between the different structures involved.
Let us point out first 
\beq
K \lop h = \deriv{}{t} \Leg^*(h) + \langle \ELform,\gam_h \rangle .
\label{K-EL}
\eeq
Here there is an abuse of notation that requires some explanation.
On the right-hand side we have a function $\Leg^*(h)$ on $\Tan Q$,
whose total time-derivative 
---see for instance
\cite{Sau-jets}
\cite{CLM-higherNoether}---
is a function on $\Tan^2Q$,
and the contraction of $\ELform$ with $\gam_h$,
considered as a function on $\Tan^2Q$;
however, 
the sum of both functions turns out to not depend on the acceleration,
so it is a function on $\Tan Q$, 
just as the left-hand side.

The local expression of (\ref{K-EL}) first appeared in
\cite{GP-lagdst}.

Though for singular lagrangians the lagrangian and the hamiltonian dynamics
are not, in general, completely determined,
equation (\ref{K-EL})
shows that, when considering solutions of 
Euler-Lagrange and Hamilton-Dirac equations,
the evolution operator $K$ gives an unambiguous time-derivative
of a function in hamiltonian space expressed in lagrangian terms.
In particular, taking $h = \phi_\mu$,
we obtain the {\it primary lagrangian constraints}
\beq
\chi_\mu := 
K \lop \phi_\mu = 
\langle \ELform,\gam_\mu \rangle 
\colon \Tan Q \to \Real ;
\label{K-ELform}
\eeq
notice that they also arise directly from (\ref{EL-ELform})
as a consistency condition
---this is due to the fact that $\gam_\mu$ are in the kernel of~$\Lleg$.
The vanishing of the primary lagrangian constraints
defines the {\it primary lagrangian subset}\/ 
$V_1 \subset \Tan Q$,
which we will assume to be a submanifold.
Notice that the functions $\chi_\mu$ are not necessarily independent,
and indeed may vanish identically.

Now we can relate the operator $K$ with the hamiltonian evolution.
A very important result for our purposes is that
\beq
K \lop h =
\Leg^*\{h,H\} +
\sum_\mu \Leg^*\{ h,\phi_\mu \} \,\lam^\mu ,
\label{K-H'}
\eeq
where there appear again the functions of equation (\ref{lam}).
The proof can be found in
\cite{BGPR-equiv},
and in
\cite{GPR-higher}
for higher-order lagrangians.
This result can be expressed also as an equality between maps
(in this case, vector fields along~$\Leg$)
rather than as an equality of differential operators:
\beq
K =
Z_H \comp \Leg +
\sum_\mu  \lam^\mu \, (Z_\mu \comp \Leg) ,
\label{K-H}
\eeq

An immediate consequence of (\ref{K-H'}) is 
\beq
\Gam_\mu \lop (K \lop h)  = \Leg^* \{ h,\phi_\mu \} .
\label{Gamma-K}
\eeq
This provides us with a test of projectability:
the function $K \lop h$ is projectable iff $h$ is a first-class function
(with respect to~$P_o$).
Recall that a function $h \colon \Tan^*Q \to \Real$ is said to be
{\it first-class}\/
with respect to a submanifold $P \subset \Tan^*Q$
if the hamiltonian vector field $Z_h$ is tangent to~$P$,
which means that
$\{h,\phi\} \feble{P} 0$
for any constraint $\phi$ defining the submanifold.
(The notation
$f \feble{M} 0$ 
means that $f(x)=0$ for all $x \in M$
(Dirac's weak equality);
for instance
$\phi_\mu \feble{P_o} 0$ and
$\chi_\mu \feble{V_1} 0$.)

\section{Some canonical vector fields}

\subsubsection*{The vector field $Y_h$}

Let $h \colon \Tan^*Q \to \Real$ be a function in phase space.
Its fibre derivative is a map
$\Der h \colon \Tan^*Q \to \Tan Q$,
so we can define another map
\beq
Y_h := \kappa \comp \Tan(\Der h) \comp K ,
\label{Yh}
\eeq
where $K$ is the time-evolution operator of~$L$
and $\kappa \colon \Tan(\Tan Q) \to \Tan(\Tan Q)$
is the canonical involution of $\Tan(\Tan Q)$.
Let us show all this in a diagram:

\indiag(3,1.05)
\capsa(0,0){\Tan Q}
\put(0,0){\LegE*{\Leg}}
\put(0,0){\LegNE{K}}

\capsa(1,0){\Tan^*Q}
\capsa(0.9,1){\Tan(\Tan^*Q)}
\put(1,1){\LegS{}}
\put(1,0){\LegE*{\Der h}}
\put(0.97,1){\LegE{\Tan(\Der h)}}

\capsa(2,0){\Tan Q}
\capsa(2,1){\Tan(\Tan Q)}
\put(2.03,1){\LegE{\kappa}}
\put(2,1){\LegS{}}

\capsa(3.06,1){\Tan(\Tan Q)}

\exdiag
\noindent
Using the local expressions of all the objects involved,
one obtains the local expression of~$Y_h$:
\beq
Y_h(q,\dot q) =
\left( 
q, \, 
\dot q; \;
\derpar{h}{p}(\Leg(q,\dot q)), \,
\dot q \dderpar{h}{q}{p}(\Leg(q,\dot q)) 
+ \derpar{L}{q} \dderpar{h}{p}{p}(\Leg(q,\dot q)) 
\right) .
\label{Yhc}
\eeq

\begin{prop}
The map $Y_h$ is a vector field on $\Tan Q$,
with local expression
\beq
Y_h = 
\Leg^*\{q,h\}   \; \derpar{}{q} + 
K \lop \{q,h\} \; \derpar{}{\dot q} .
\label{Yhcc}
\eeq 
It has the following properties:
\bea
&&
\vend \comp Y_h = \Gam_h ,
\\
&&
Y_g \lop (\Leg^* h)   = \Leg^* \{h, g \} + \Gam_h \lop (K \lop g) ,
\label{Y-Leg}
\\
&&
Y_g \lop (K \lop h) = K \lop \{h, g\} + Y_h \lop (K \lop g) ,
\label{Y-K}
\\
&&
\Tan(\Leg) \comp Y_g = Z_g \comp \Leg + \Ups^{K \lop g} .
\label{Leg-Y}
\eea
\end{prop}

\proof
The fact that $Y_h$ is a vector field is a direct consequence of 
its local expression (\ref{Yhc}).
It follows also from
$$
\tau_{\Tan Q} \comp Y_h = 
\tau_{\Tan Q} \comp \kappa \comp \Tan(\Der h) \comp K =
\Tan(\tau_Q) \comp \Tan(\Der h) \comp K =
\Tan(\tau_Q^*) \comp K =
\Id_{\Tan Q} .
$$

The alternative (and more suggestive) local expression 
(\ref{Yhcc}) of $Y_h$ is also clear from (\ref{Yhc}),
as well as the fact that $\vend \comp Y_h = \Gam_h$
---$\vend$ is the vertical endomorphism of $\Tan(\Tan Q)$.

The following two equations can be proved from their local expressions.
This is simpler for the first one, (\ref{Y-Leg}):
its left and right-hand sides read in coordinates
$$
\left(
 \widehat{\derpar{h}{q}} +
 \widehat{\derpar{h}{p}} \dderpar{L}{\dot q}{q}
\right) 
\widehat{\derpar{g}{p}}
+
\widehat{\derpar{h}{p}}
\dderpar{L}{\dot q}{\dot q}
\left(
 \widehat{\dderpar{g}{p}{q}} \dot q +
 \widehat{\dderpar{g}{p}{p}} \derpar{L}{q} 
\right) 
$$
(we have put $\widehat h = \Leg^*h$ to simplify the notation).

Regarding the second equation, (\ref{Y-K}),
one has to prove
$Y_g \lop (K \lop h) - Y_h \lop (K \lop g) = K \lop \{h, g \}$.
The terms remaining after 
the antisymmetrisation of $Y_g (K \lop h)$ with respect to $(g,h)$
can be arranged to read
$$
\left(
 \dot q \;        \Leg^* \; \derpar{}{q} +
 \derpar{L}{q} \; \Leg^* \; \derpar{}{p} 
\right)
\left(
 \derpar{h}{q} \derpar{g}{p} -
 \derpar{h}{p} \derpar{g}{q}
\right) ,
$$
which is $K \lop \{h, g\}$.

Finally, (\ref{Leg-Y}) is obtained by using relation (\ref{Ups-Gam})
to express equation (\ref{Y-Leg}) as an equality 
between vector fields along~$\Leg$.
\qed

\subsubsection*{The vector fields $R_h$ and $\Del_h$}

Equation (\ref{Leg-Y}) shows explicitly an obstruction for 
the projectability of $Y_g$ to the hamiltonian vector field~$Z_g$.
In the discussion of this issue it will be interesting to consider
the vertical vector field
\beq  
R_h = 
\Gam_{\{h,H\}} + \lam^\mu \Gam_{\{h,\phi_\mu\}} ,
\label{R}
\eeq
defined from any function $h$ on phase space
---from now on we use the summation convention
for the greek indices associated with the primary constraints.
Notice that $R_h$ depends on the choice of the hamiltonian $H$
and the primary hamiltonian constraints~$\phi_\mu$.
The action of $R_h$ on projectable functions is
\beq
R_g \lop \Leg^*h = 
\Gam_h \lop (K \lop g) - \Leg^*\{g,\phi_\mu\} \, \Gam_h \lop \lam^\mu ,
\label{R-Leg}
\eeq
which is a kind of generalisation of (\ref{Gamma-K}).
To prove it,
first we apply $R_g$ to $\Leg^*h$, 
then we use the symmetry property
\beq
\Gam_h \lop \Leg^*(g) = 
\Lleg \opers{\gam_g,\gam_h} =
\Gam_g \lop \Leg^*(h) ,
\label{Wsim}
\eeq
and finally we apply equation (\ref{K-H'}) to let $K$ appear explicitly.

\medskip

The interest of the vector field $R_h$ comes from the fact that
it appears when taking equation (\ref{Y-K}) 
and rewriting it using relations 
(\ref{K-H'}) and (\ref{Y-Leg});
after some cancellations one arrives at
\beq
R_h \lop( K \lop g ) + \Leg^*\{ h,\phi_\mu \} \,Y_g \lop \lam^\mu = 
R_g \lop( K \lop h ) + \Leg^*\{ g,\phi_\mu \} \,Y_h \lop \lam^\mu .
\label{newop}
\eeq
In other words, the left-hand side is symmetric in $(g,h)$.
We can develop this further, applying equation (\ref{K-H'}) again
to make $K$ disappear from (\ref{newop}).
A convenient organisation of the terms,
together with some additional cancellations due to 
the symmetry property (\ref{Wsim}),
finally yields another symmetric equation:
\beq
\Leg^*\{ h,\phi_\mu \} \; (Y_g - R_g) \lop \lam^\mu =  
\Leg^*\{ g,\phi_\mu \} \; (Y_h - R_h) \lop \lam^\mu .
\label{Delta-lam-previ}
\eeq

This suggests to define,
for any function $g$ in phase space, 
the vector field
\beq  
\Del_g = Y_g - R_g .
\label{Delta}
\eeq

\begin{prop}
\label{prop-Delta}
The vector field $\Del_g$ has the following properties:
\bea
&&
\vend \comp \Del_g = \Gam_g ,
\label{J-Delta}
\\
&&
\Del_g \lop \lam^\mu = 
- \Leg^*\{g,\phi_\nu\} \, M \opers{\Der \lam^\mu,\Der \lam^\nu} ,
\label{Delta-lam}
\\
&&
\Del_g \lop (\Leg^*h) = 
\Leg^*\{h,g\} + \Leg^*\{g,\phi_\mu\} \, \Gam_h \lop \lam^\mu ,
\label{Delta-Leg}
\\
&&
\Tan(\Leg) \comp \Del_g = 
Z_g \comp \Leg + \Leg^*\{g,\phi_\mu\} \, \Ups^{\lam^\mu} .
\label{Leg-Delta}
\eea
\end{prop}

\proof
The first property is a consequence of the same property of~$Y_g$
and the fact that $R_g$ is vertical.

The second property gives the action of $\Del_g$
on the non-projectable functions~$\lam^\mu$.
To prove it, we consider equation (\ref{Delta-lam-previ}),
$$
\Leg^*\{ h,\phi_\mu \} \; \Del_g \lop \lam^\mu =  
\Leg^*\{ g,\phi_\mu \} \; \Del_h \lop \lam^\mu ;
$$
taking for $h$ the configuration variables $h=q^i$, one gets 
$$
(\Del_g \lop \lam^\mu) \, \gam_\mu =
- \Leg^*\{g,\phi_\mu\} \, M \scomp \Der \lam^\mu ,
$$
with $M \colon \Tan Q \to \Hom(\Tan^*Q,\Tan Q)$
given by equation (\ref{M}).
Then contraction with $\Der \lam^\nu$
and use of the property (\ref{lam-gam}) finally yields
equation (\ref{Delta-lam}).

Subtracting equations (\ref{Y-Leg}) and (\ref{R-Leg}) 
yields (\ref{Delta-Leg}).

Finally, using the relation (\ref{Ups-Gam})
we can remove the function $h$ from the preceding equation
to obtain an equality between vector fields along~$\Leg$,
thus obtaining (\ref{Leg-Delta}).
\qed

\subsubsection*{Some additional properties}

The vector field on $\Tan Q$ $\Gam_h$ and 
the vector field along $\Leg$ $\Ups^f$
are defined in terms of the fibre derivative,
and a trivial application of Leibniz's rule shows that
\bea
&&
\Gam_{h_1h_2} =
\Leg^*(h_1) \Gam_{h_2} + \Leg^*(h_2) \Gam_{h_1} ,
\\
&& 
\Ups^{f_1f_2} = f_1 \Ups^{f_2} + f_2 \Ups^{f_1} .
\eea
Similarly one can compute
\bea
&& 
Y_{h_1h_2} =
\Leg^*(h_1) Y_{h_2} + \Leg^*(h_2) Y_{h_1} +
(K \lop h_1) \Gam_{h_2} + (K \lop h_2) \Gam_{h_1} ,
\\
&& 
R_{h_1h_2} =
\Leg^*(h_1) R_{h_2} + \Leg^*(h_2) R_{h_1} +
(K \lop h_1) \Gam_{h_2} + (K \lop h_2) \Gam_{h_1} ,
\\
&&
\Del_{h_1h_2} =
\Leg^*(h_1) \Del_{h_2} + \Leg^*(h_2) \Del_{h_1} .
\eea
The last equation, 
which is obtained immediately by subtracting the two previous ones,
shows that the vector field $\Del_h$ is also 
a first-order differential operator on~$h$.

\section{Applications to the kinematics}

\subsubsection*{The projectability to a hamiltonian vector field}

In equations
(\ref{Delta-lam}), (\ref{Delta-Leg}) and (\ref{Leg-Delta})
there is a common piece $\Leg^*\{g,\phi_\mu\}$
whose vanishing gives an answer to the question of projectability:

\begin{teor}
Let $L$ be an almost regular lagrangian.
The necessary and sufficient condition 
for the hamiltonian vector field $Z_g$ in $\Tan^*Q$ 
to be the projection (through the Legendre transformation)
of a vector field in $\Tan Q$ is that
$g$ should be a first-class function with respect to 
the primary hamiltonian constraint submanifold $P_o \subset \Tan^*Q$.

Then the vector field $\Del_g$ projects to~$Z_g$:
\beq
\Tan(\Leg) \comp \Del_g  = 
Z_g \comp \Leg  .
\label{Leg-Delta'}
\eeq
Any other vector field projecting to~$Z_g$ is obtained by
adding to $\Del_g$ 
any vector field in the kernel of the tangent map $\Tan(\Leg)$. 
\end{teor}

\proof
As we have said in section~2,
the condition for a vector field in $\Tan^*Q$ to be a projection
is its tangency to~$P_o = \Leg(\Tan Q)$.
When this vector field is the hamiltonian vector field~$Z_g$ 
this means that $g$ is a first-class function
with respect to the primary constraint submanifold~$P_o$,
that is,
$\Leg^*\{g,\phi_\mu\}=0$.
Then (\ref{Leg-Delta}) shows that $\Del_g$ projects to~$Z_g$.

The last assertion is obvious,
since the vector fields that project to zero are those in $\Ker \Tan(\Leg)$.
\qed

\medskip

Comparing (\ref{Leg-Delta}) and (\ref{Leg-Y}) one realises that
the appropriate vector field candidate to project to $Z_g$ is $\Del_g$.
This is because the condition that $\Ups^{K \lop g}=0$,
which is equivalent to $\Der(K \lop g) = 0$,
is more restrictive than $g$ being first-class.
Indeed, $\Der(K \lop g) = 0$ means that any vertical vector field 
acting on $K \lop g$ yields zero, then in particular 
$\Gam_\mu \lop (K \lop g)  = \Leg^* \{ g,\phi_\mu \} = 0$
by (\ref{Gamma-K}).
Of course, when $\Der(K \lop g) = 0$
we can say that also $Y_g$ projects to~$Z_g$.
This is also a consequence of the fact that
if $\Der(K \lop g) = 0$ then $R_g$ is in $\Ker \Tan(\Leg)$.

Equation (\ref{Leg-Delta'}) in the theorem 
is a direct consequence of 
equation (\ref{Leg-Delta}) in proposition~\ref{prop-Delta}
when $g$ is first-class.
Let us rewrite equations 
(\ref{Delta-lam}) and (\ref{Delta-Leg}) accordingly:

\begin{prop}
Let $g\colon \Tan^*Q \to \Real$ be a first-class function
with respect to the primary hamiltonian constraint submanifold 
$P_o \subset \Tan^*Q $.
Then the following results hold:
\bea
&&
\Del_g \lop \lam^\mu = 0 ,
\label{Delta-lam'}
\\
&&
\Del_g \lop \Leg^*h =  \Leg^* \{h,g\}
\hbox{ for any function~$h$.}
\label{Delta-Leg'}
\eea
\qed
\end{prop}

\vskip -3mm
Recalling (\ref{lam-gam}),
$\Gam_\nu \lop \lam^\mu 
= \delta^{\mu}_{\,\nu}$,
notice that equation (\ref{Delta-lam'}) singles out $\Del_g$,
among the set of vector fields projecting to~$Z_g$,
as the only one
whose action on the non-projectable functions $\lam^\mu$ is zero.

Now let us study some commutators among vector fields:

\begin{prop}
Let $\phi, \phi' \colon \Tan^*Q \to \Real$ 
be primary hamiltonian constraints,
and $g, g' \colon \Tan^*Q \to \Real$
be first-class functions 
with respect to the primary hamiltonian constraint submanifold 
$P_o \subset \Tan^*Q $.
Then the following results hold:
\bea
&&
[\Gam_\phi,\Gam_{\phi'}] = 0 ,
\label{com-Gam-Gam}
\\
&&
[\Del_g,\Del_{g'}] = -\Del_{\{g,g'\}} ,
\label{com-Del-Del}
\\
&&
[\Del_g,\Gam_\phi] = -\Gam_{\{g,\phi\}} - [R_g-\Gam_{\{g,H\}},\Gam_\phi] .
\label{com-Del-Gam}
\eea
\label{commuts}
\end{prop}
\vskip -2mm

\proof
The first result is well known, we include it for the sake of
completeness,
and it is readily proved in coordinates
taking into account that $\Gam_\phi \lop \Leg^*(h) = 0$
for any function~$h$.

For the second result, 
to show the equality of both vector fields
it is enough to prove that both coincide 
as differential operators when acting 
on projectable functions 
(this is a consequence of equation (\ref{Delta-Leg'}),
together with $[Z_g,Z_{g'}] = Z_{\{g',g\}}$)
and on the non-projectable functions~$\lam^\mu$
(this is a trivial consequence of equation (\ref{Delta-lam'})).

One can proceed in the same way to prove the third commutator.
To this end, we first prove that
\beq
[\Del_g,\Gam_\mu] = 0 .
\label{com-Del-mu}
\eeq
On projectable functions the Lie bracket of the vector fields is zero;
this is due to equation (\ref{Delta-Leg'}), 
and the fact that $\Gam_\mu$ applied
to any projectable function gives zero.
On the non-projectable functions $\lam^\mu$,
equation (\ref{Delta-lam'}) 
and the fact that $\Gam_\mu \lop \lam^\nu$ is constant
also yields zero.

Now let us deal with the general case.
First, locally we can express $\phi = a^\mu \phi_\mu$
for some functions~$a^\mu$.
Then
$$
\Gam_{a^\mu \phi_\mu} =
\Leg^*(a^\mu) \Gam_\mu$$ 
and
$
[\Del_g,\Gam_\phi] =
[\Del_g,\Leg^*(a^\mu) \Gam_\mu] =
\Del_g \lop \Leg^*(a^\mu) \, \Gam_\mu ,
$
thanks to (\ref{com-Del-mu}).
Using (\ref{Delta-Leg'}) we obtain
$$
[\Del_g,\Gam_\phi] =
\Leg^*\{a^\mu,g\} \, \Gam_\mu .
$$
Considering $\{g,\phi\}$ we have
$
\Gam_{\{g,\phi\}} =
\Leg^*(a^\mu) \Gam_{\{g,\phi_\mu\}} +
\Leg^*\{g,a^\mu\} \Gam_{\mu}
$, and so we get
$$
[\Del_g,\Gam_\phi] + \Gam_{\{g,\phi\}} =
\Leg^*(a^\mu) \Gam_{\{g,\phi_\mu\}} .
$$
Finally,
$\Gam_\phi \lop \lam^\mu = \Leg^*(a^\mu)$,
so we arrive at
\beq
[\Del_g,\Gam_\phi] + \Gam_{\{g,\phi\}} =
(\Gam_\phi \lop \lam^\mu) \, \Gam_{\{g,\phi_\mu\}} .
\label{com-Del-Gam'}
\eeq
To obtain (\ref{com-Del-Gam}),
notice that by definition
$R_g - \Gam_{\{g,H\}} = \lam^\mu \Gam_{\{g,\phi_\mu\}}$,
and since by (\ref{com-Gam-Gam}) 
the $\Gam$'s of constraints commute,
$
[R_g-\Gam_{\{g,H\}},\Gam_\phi] = 
[\lam^\mu \Gam_{\{g,\phi_\mu\}},\Gam_\phi] =
- (\Gam_\phi \lop \lam^\mu) \, \Gam_{\{g,\phi_\mu\}}
$.
\qed

Notice moreover that using the relation between $Y_g$ and $\Del_g$
we can rewrite equation (\ref{com-Del-Gam}) as
\beq
[\Del_g + \lam^\mu \Gam_{\{g,\phi_\mu\}} , \Gam_\phi] = 
\Gam_{\{\phi,g\}} =
[Y_g-\Gam_{\{g,H\}},\Gam_\phi] .
\eeq

\subsubsection*{The kernel of the presymplectic form in $\Tan Q$}

Here we will show that the vector fields $\Del_g$
provide an easy explicit construction of the kernel of
the presymplectic form
$\omega_L = \Leg^* \omega_Q$
of the lagrangian formalism.

If a vector field $Y$ in $\Tan Q$ projects through $\Leg$ 
to a vector field $Z$ in $\Tan^*Q$, we have
$$
i_Y \,\omega_L = \Leg^* \left( i_Z \,\omega_Q \right) .
$$ 
This shows trivially that 
$\Ker \Tan(\Leg) \subset \Ker \omega_L$
---indeed it is a well-known fact that
$\Ker \Tan(\Leg) = \Ker \omega_L \cap \Ver(\Tan Q)$.
So the vector fields $\Gam_\mu$ are part of a basis for $\Ker \omega_L$.

Now let us assume that
the matrix of Poisson's brackets
$\{\phi_\mu,\phi_\nu\}$
has constant rank.
Then one can find an appropriate set $(\phi_\mu)$
of independent primary hamiltonian constraints
which are split into first-class $\phi_{\mu_o}$ 
---their Poisson bracket with any primary hamiltonian constraint
vanishes on~$P_o$---
and second-class $\phi_{\mu_o'}$
---see among others
\cite{DLGP-HJ}.
Being the functions $\phi_{\mu_o}$ first-class,
the corresponding vector field $\Del_{\mu_o} = \Del_{\phi_{\mu_o}}$ projects to
the hamiltonian vector field $Z_{\mu_o}$, and since
$$
i_{\Del_{\mu_o}} \omega_L =
\Leg^*\left( i_{Z_{\mu_o}} \omega_Q \right) = 
\Leg^* ( \dif \phi_{\mu_o} ) = 
\dif \Leg^*(\phi_{\mu_o}) = 0 ,
$$ 
we conclude that $\Del_{\mu_o}$ is also in $\Ker \omega_L$.

Notice that the vector fields $\Del_\mu$ are linearly independent,
since application of the vertical endomorphism 
yields independent vector fields, 
$\vend \comp \Del_\mu = \Gam_\mu$;
moreover, they are also independent of the~$\Gam_\mu$.
Finally, the dimension of $\Ker \omega_L$ and the number of
primary hamiltonian constraints plus the number of the first-class ones
coincide
---see for instance
\cite{MMS-constraints}.
So we have proved the following result:

\begin{teor}
The kernel of $\omega_L$ has a basis constituted by
the vector fields $\Gam_\mu$, 
associated with the primary hamiltonian constraints $\phi_\mu$,
and the vector fields $\Del_{\mu_o}$, 
associated with a basis of 
the first-class primary hamiltonian constraints $\phi_{\mu_o}$.
\qed
\end{teor}

This kernel has been studied in the literature on singular lagrangians
for its interest in the classification of the constraints
\cite{CLR-origin}
\cite{Car-theory}
\cite{MR-lag}.
An explicit computation of the kernel was first presented in 
\cite{PSS-quot99} 
(see equations (2.13a) and (2.13b) of that paper),
but in a coordinate, rather than geometric, framework.
In that paper the kernel was given in a slightly different basis, 
for $\Del_{\mu_o}$ in that paper is the present $\Del_{\mu_o}$ 
except for the term $\lam^\nu \Gam_{\{\phi_{\mu_o},\phi_\nu\}}$, 
which is a combination of the vector fields $\Gam_\mu$, 
also in the kernel. 
The present basis is preferable
because it gives the commutation relations in their simplest form. 
Indeed, if
$$
\{ \phi_{\mu_o},\phi_{\nu_o} \} =  
 B_{{\mu_o}{\nu_o}}^{\rho_o}\phi_{\rho_o} 
+ O(\phi^2),
$$
(the Poisson's bracket of first-class constraints is first-class), 
then, taking into account proposition~\ref{commuts},
the algebra reads
\bea
[\Gam_{\mu},    \Gam_{\nu}]     &=& 0 , \nonumber \\{}
[\Gam_{\mu},    \Del_{\nu_o}] &=& 0 , \\{}
[\Del_{\mu_o},\Del_{\nu_o}] &=& \Leg^*(B_{{\nu_o}{\mu_o}}^{\rho_o})
                                    \,\Del_{\rho_o} \nonumber.
\eea

\section{Applications to dynamics and symmetries}

\subsubsection*{Lagrangian dynamics}

Here we will give an explicit expression of the lagrangian dynamics
in terms of vector fields.
Though in the case of a singular lagrangian 
the Euler-Lagrange equation can not be written in normal form,
one can try to express its solutions in terms of integral curves
of some dynamical vector fields.
For instance,
consider the Euler-Lagrange equation in the form
(\ref{EL-K}):
$\Tan(\Leg) \comp \dot\xi = K \comp \xi$.
Let $V \subset \Tan Q$ be a submanifold and
$\XL$ a second-order vector field in $\Tan Q$
{\it tangent}\/ to~$V$.
Then the integral curves of $\XL$ contained in $V$ are solutions of the
Euler-Lagrange equation iff $\XL$ satisfies
\beq
\Tan(\Leg) \comp \XL \feble{V} K ,
\eeq
(the weak equality means equality on the points of the submanifold~$V$).

As a first approximation to this problem,
let us call $V_1$ the subset of points $u \in \Tan Q$ where the linear equation
---for the unknown vector~$a_u$---
$\Tan_u(\Leg) \lop a_u = K(u)$
is consistent, and assume it to be a submanifold,
the primary lagrangian constraint submanifold.
Then the equation
\beq
\Tan(\Leg) \comp \XL \feble{V_1} K 
\label{EL-X-prim}
\eeq
has solutions, let us call them {\it primary dynamical vector fields}
\cite{GP-gener}.
They are not unique on~$V_1$,
since they can be added vector fields in $\Ker \Tan(\Leg)$.
On the other hand, one should find solutions that are tangent to~$V_1$,
and this is the beginning of an algorithm that,
under some regularity conditions,
may give at the end all the solutions of the Euler-Lagrange equation.
This is like the Dirac's theory in lagrangian formalism
---see a careful discussion in
\cite{GP-gener};
see also
\cite{BGPR-equiv}
\cite{MR-lag}.

Notice that any integral curve of a primary dynamical field~$\XL$
which is {\it contained}\/ in~$V_1$
is a solution of the Euler-Lagrange equation.

\medskip

Our purpose now is to show that
the choice of the hamiltonian function $H$ 
and the set of primary hamiltonian constraints~$\phi_\mu$
yields a primary dynamical~field $\XL$.
Let us define the vector field
\beq
\XL_o = \Del_H + \lam^\mu \Del_\mu .
\label{XL}
\eeq

\begin{teor}
The vector field $\XL_o $
satisfies the second-order condition,
and is a primary dynamical field.
More precisely,
\beq
\Tan(\Leg) \comp \XL_o 
= K - \chi_\mu \Ups^{\lam^\mu} 
\feble{V_1} K .
\label{K-XL}
\eeq
\end{teor}

\proof
A second-order vector field on $\Tan Q$ can be characterised by
the property that $\vend \comp X = \Lio_{\Tan Q}$.
We have
$$
\vend \comp (\Del_H + \lam^\mu \Del_\mu) =
\Gam_H + \lam^\mu \Gam_\mu =
\Lio_{\Tan Q} ,
$$
by (\ref{J-Delta}) and (\ref{lam'}),
so $\XL_o$ satisfies the second-order condition.

Now let us apply $\Tan(\Leg)$ to $\XL_o$, and use (\ref{Leg-Y}):
$$
\Tan(\Leg) \comp \XL_o = 
Z_H \comp \Leg + \lam^\mu Z_\mu \comp \Leg +
\left(\strut
 \Leg^*\{H,\phi_\mu\} + \lam^\nu\, \Leg^*\{\phi_\nu,\phi_\mu\} 
\right) \; \Ups^{\lam^\mu} .
$$
In this expression we recognise the operator~$K$
---see equation (\ref{K-H'})--- 
and the primary lagrangian constraints $\chi_\mu = K \lop \phi_\mu$,
thus obtaining (\ref{K-XL}).
\qed

Before proceeding it will be interesting to notice some additional
properties of~$\XL_o$.
(We will use the notation 
$Y_\mu = Y_{\phi_\mu}$ and $R_\mu = R_{\phi_\mu}$.)

\begin{prop}
The vector field $\XL_o$ satisfies the following properties:
\bea
\XL_o &=& Y_H + \lam^\mu Y_\mu ,
\\
\XL_o \lop \Leg^*(h) &=& K \lop h - \chi_\mu \,\Gam_h \lop \lam^\mu ,
\label{XL-Leg}
\\
\XL_o \lop \lam^\nu &=&
\chi_\mu \, M \opers{\Der \lam^\nu, \Der \lam^\mu} 
\feble{V_1} 0 ,
\label{XL-lam}
\\
\XL_o \lop (K \lop h) &=&
K \lop \{h,H\} + \lam^\mu K \lop \{h,\phi_\mu\} +
\nonumber 
\\
&& +
\chi_\nu \left(\strut
  - R_h \lop \lam^\nu +
  \Leg^*\{h,\phi_\mu\} M \opers{\Der \lam^\mu, \Der \lam^\nu} 
\right) .
\label{XL-K}
\eea
\end{prop}

\proof
The first statement is an immediate consequence of the definition of
$\XL_o$ and the fact that
\beq
R_H + \lam^\nu R_\nu = 0 ,
\eeq
whose proof is
$
R_H + \lam^\nu R_\nu =
- \lam^\mu \Gam_{\{\phi_\mu,H\}} 
+ \lam^\nu \left(
    \Gam_{\{\phi_\nu,H\}} + \lam^\mu \Gam_{\{\phi_\nu,\phi_\mu\}}
  \right) =
\Gam_{\{\phi_\nu,\phi_\mu\}} \lam^\nu \lam^\mu = 
0 
$,
due to the antisymmetry of $\{\phi_\nu,\phi_\mu\}$.

The second one is a direct consequence of equation (\ref{K-XL}):
it tells us the action of $\XL_o$ 
(and indeed of any primary dynamical field $\XL$) 
on projectable functions.

The third equation gives the action of $\XL_o$ 
on the non-projectable functions~$\lam^\mu$.
It is obtained from (\ref{Delta-lam})
and the definition of the primary lagrangian constraints $\chi_\mu$:
\beann
\XL_o \lop \lam^\nu 
&=&
(\Del_H + \lam^\mu \Del_\mu) \lop \lam^\nu =
\left( \Leg^*\{\phi_\mu,H\} + \lam^\rho \Leg^*\{\phi_\mu,\phi_\rho\} \right)
\, M \opers{\Der \lam^\nu, \Der \lam^\mu} 
\\
&=&
K \lop \phi_\mu \, M \opers{\Der \lam^\nu, \Der \lam^\mu} =
\chi_\mu        \, M \opers{\Der \lam^\nu, \Der \lam^\mu} .
\eeann

The fourth equation is obtained from
$K \lop h =
\Leg^*\{h,H\} +
\sum_\mu \Leg^*\{ h,\phi_\mu \} \,\lam^\mu 
$,
(\ref{K-H'}),
by applying
(\ref{XL-Leg}) and (\ref{XL-lam}).
\qed

As a consequence of the theorem we obtain the general form of 
a primary dynamical field in lagrangian formalism:
$$
\XL = \XL_o + \eps^\mu \,\Gam_\mu .
$$
On the other hand, according to (\ref{HD}),
the primary dynamical fields
in hamiltonian formalism are
$$
\XH = Z_H + \lambda^\mu \,Z_\mu .
$$
Both vector fields exhibit a set of arbitrary functions,
$\eps^\mu$ on $\Tan Q$ and $\lambda^\mu$ on $\Tan^*Q$,
and we can relate the corresponding dynamics:

\begin{prop}
Let $\xi \colon I \to \Tan Q$, $\eta \colon I \to \Tan^*Q$
related solutions of the Euler-Lagrange and Hamilton-Dirac equations
corresponding to the dynamical vector fields
$$
\XL = \XL_o + \eps^\mu \,\Gam_\mu ,
\qquad
\XH = Z_H + \lambda^\mu \,Z_\mu .
$$
Then the ``arbitrary functions'' $\eps^\mu$, $\lambda^\mu$
are related by
\bea
\lambda^\mu(\eta(t)) &=& \lam^\mu(\xi(t)) ,
\label{arb-ham}
\\
\eps^\mu(\xi(t)) &=& (K \lop \lambda^\mu)(\xi(t)) .
\label{arb-lag}
\eea
\end{prop}

\proof
We have
$$
\dot\eta = Z_H \comp \eta + (\lambda^\mu \comp \eta) \, Z_\mu \comp \eta .
$$
Since $\xi$ and $\eta$ are related,
application of $\Tan(\tau_Q^*)$ yields
$$
\xi = 
\Der H \comp \eta + (\lambda^\mu \comp \eta) \, \Der \phi_\mu \comp \eta =
\Der H \comp \Leg \comp \xi + 
(\lambda^\mu \comp \eta) \, \Der \phi_\mu \comp \Leg \comp \xi ,
$$
and from (\ref{lam})
$$
\xi = 
\gam_H \comp \xi + (\lam^\mu \comp \xi) \, \gam_\mu \comp \xi ;
$$
comparing both expressions we identify $\lambda^\mu$ with $\lam^\mu$.

Now we compute
\beann
(K \lop \lambda^\mu)(\xi(t)) &=&
\deriv{}{t} \lambda^\mu(\eta(t)) =
\deriv{}{t} \lam^\mu(\xi(t)) 
\\
&=&
\XL \lop \lam^\mu =
(\XL_o + \eps^\nu \,\Gam_\nu) \lop \lam^\mu 
\\
&=&
\eps^\mu(\xi(t)) ,
\eeann
where we have used (\ref{arb-ham}) and the properties
$\XL_o \lop \lam^\mu \feble{V_1} 0$, 
$\Gam_\nu \lop \lam^\mu = \delta^\mu_\nu$.
\qed

Another application of the properties of $\XL_o$
is the relation between the lagrangian and the hamiltonian 
constraint algorithms.
For instance, putting $\phi_\mu^1 = \{\phi_\mu,H\}$
---this is a secondary hamiltonian constraint when $\phi_\mu$
is first-class---, 
from (\ref{XL-K}) we have
$$
\XL_o \lop (K \lop \phi_\rho) =
K \lop \phi_\rho^1 + \lam^\mu K \lop \{\phi_\rho,\phi_\mu\} +
\chi_\nu \left(\strut
  - R_\rho \lop \lam^\nu +
  \Leg^*\{\phi_\rho,\phi_\mu\} M \opers{\Der \lam^\mu, \Der \lam^\nu} 
\right) ,
$$
and so for first-class constraints we get
$$
\XL_o \lop (K \lop \phi_{\mu_o}) \feble{V_1} K \lop \phi_{\mu_o}^1 ,
$$
which means that 
performing the first step of the hamiltonian stabilisation 
followed by application of~$K$
is equivalent to
applying $K$ and then 
performing the first step of the lagrangian stabilisation. 

In a similar way from (\ref{XL-Leg}) we obtain
$$
\XL_o \lop \Leg^*\phi_{\mu_o}^1 \feble{V_1} K \lop \phi_{\mu_o}^1 .
$$

\smallskip

In 
\cite{BGPR-equiv}
a vector field similar to the dynamical vector field $\XL_o$ 
was introduced in coordinates,
and was used in 
\cite{Pon-newrel}
to explore the relations between lagrangian and hamiltonian dynamics
for singular lagrangians.
However, the simplest way to relate both dynamics is achieved
with the choice of~$\XL_o$.

On the other hand, in 
\cite{Gra-fibder}
an intrinsic way to construct a primary dynamical field
in lagrangian formalism out from any second-order vector field
was introduced
using the Euler-Lagrange operator $\ELform$ 
and the map $M$ given by equation (\ref{M}).
This procedure,
when applied to the primary dynamical fields,
leaves them invariant ``on-shell'' 
(we mean on the primary lagrangian constraint submanifold).
The vector field $\XL_o$ is special among the primary dynamical fields
in the sense that
its action on the non-projectable functions $\lam^\mu$ is zero on-shell.

\subsubsection*{Canonical symmetries and canonical Noether symmetries}
 
Now we will re-express some statements about symmetries
using the vector field~$Y_h$.

Let us consider the time-independent symmetries in phase space that are
generated by a function $G$ on phase space
through the hamiltonian vector field 
$Z_G = \{-,G\}$. 
It turns out 
\cite{GP-hamdst} 
that the necessary and sufficient condition for a function $G$ 
to generate in this way 
an infinitesimal symmetry of the Hamilton-Dirac equation of motion is that 
\beq
K \lop G \forta{V_f} c,
\label{KGeq}
\eeq 
for some constant~$c$
(in the time-dependent case this would be a function $c(t)$).
Here $\forta{}$ stands for Dirac's strong equality, that is, 
an equality up to quadratic terms in the constraints 
---now the whole set of constraints,
corresponding to the final lagrangian constraint submanifold~$V_f$
\cite{BGPR-equiv}
\cite{GP-gener}. 

Then, application of (\ref{Y-K}) yields
\beq
Y_G \lop (K \lop h)  \feble{V_f}  K \lop \{h, G \} 
\label{YG-Kh}
\eeq
for every function~$h$,
where $\feble{}$ means equality on the whole constraint surface. 

Notice conversely that if a function $G$ satisfies (\ref{YG-Kh}) 
for every function~$h$, then (\ref{Y-K}) implies that
$
Y_h \lop (K \lop G) \feble{V_f} 0
$
for each~$h$,
and so we obtain (\ref{KGeq}) again.
We have thus obtained the following:

\begin{teor}
The necessary and sufficient condition for the hamiltonian vector field
$Z_G$ to generate
a symmetry of the Hamilton-Dirac equation of motion is
\beq
Y_G \lop (K \lop h) \feble{V_f} K \lop (Z_G \lop h)
\label{weakcomm}
\eeq
for all functions~$h$.
\qed
\end{teor}

One can also consider the more restrictive case of 
canonical Noether symmetries,
whose infinitesimal generator $G$ can be characterised in a similar way 
\cite{BGGP-noether}
as
\beq
K \lop G = c .
\label{KGequal}
\eeq 
Then the same reasoning as above leads to the following:

\begin{teor}
The necessary and sufficient condition for
the hamiltonian vector field $Z_G$ to generate
a Noether symmetry in phase space is that 
\beq
Y_G \lop (K \lop h) = K \lop (Z_G \lop h)
\label{goodcomm}
\eeq
for all functions~$h$.
\qed
\end{teor}

Notice the remarkable fact that a weak (on-shell) equality 
or a standard equality 
is the only difference between 
the characterisation (\ref{weakcomm}) for a symmetry of 
the Hamilton-Dirac equation of motion and  
the characterisation (\ref{goodcomm}) for a canonical Noether symmetry. 
Since Noether symmetries exhibit a property of the action functional, 
it is clear that their characterisation must be, as we see, 
on-shell and off-shell. 
This characterisation (\ref{goodcomm}) was first obtained 
in the paper 
\cite{GP-commute}, 
which was instrumental in finding the new geometric structures 
that have been introduced in the present paper.

Notice also that, when $c \neq 0$ in (\ref{KGeq}) or (\ref{KGequal}),
the conserved quantity associated to the symmetry is
$G-ct$ rather than~$G$.

\section{The case of a regular lagrangian}

In this section we will show what the preceding results become
when the lagrangian is hyperregular,
namely, when $\Leg \colon \Tan Q \to \Tan^*Q$
is a diffeomorphism
---in a local study,
we might suppose only that the lagrangian is regular,
namely, that $\Leg$ is a local diffeomorphism.

Now the 2-form $\omega_L = \Leg^*(\omega_Q)$ on $\Tan Q$ is symplectic.
Let us denote by $X_f$ the hamiltonian vector field of a function $f$
with respect to~$\omega_L$.
Recall that the lagrangian dynamics is now ruled by 
the hamiltonian vector field $\XL = X_{\En}$ of the energy function.

\begin{prop}
Suppose that the lagrangian is hyperregular. Then:
\bea
&&
\Gam_h = \vend \comp X_{\Leg^*(h)} ,
\label{Gam-reg}
\\
&&
R_h = \vend \comp X_{\Leg^*\{h,H\}} ,
\label{R-reg}
\\
&&
\Del_h = X_{\Leg^*h} ,
\label{Delta-reg}
\\
&&
Y_h = X_{\Leg^*(h)} + \vend \comp X_{\Leg^*\{h,H\}} .
\label{Y-reg}
\eea
\end{prop}
\proof
The vertical vector fields in (\ref{Gam-reg})
correspond to bundle maps $\Tan Q \to \Tan Q$.
For the right-hand side the map is
$$
\Tan(\tau_Q) \comp X_{\Leg^*(h)} = 
\Tan(\tau_Q) \comp \Tan(\Leg^{-1}) \comp Z_h \comp \Leg = 
\Tan(\tau_Q^*) \comp Z_h \comp \Leg
$$
which coincides with 
the map $\gam_h = \Der h \comp \Leg$ that corresponds to~$\Gam_h$.

Definition (\ref{R}) when there are no constraints yields
$R_h = \Gam_{\{h,H\}}$.
Then equation (\ref{R-reg}) follows immediately from (\ref{Gam-reg}).
(Notice by the way that $R_H = 0$.)

Another consequence of the non existence of constraints is that,
according to (\ref{Leg-Delta}) or theorem~1,
$\Del_h$ projects to the hamiltonian vector field~$Z_h$,
and thus it is the hamiltonian vector field of $\Leg^*(h)$,
which is the contents of (\ref{Delta-reg}).

Finally, the last equation is an immediate consequence of the definition 
$\Del_h = Y_h-R_h$.
\qed

\smallskip
Given a second-order vector field $D$ on $\Tan Q$,
a vector field $X$ is called {\it newtonoid}\/ with respect to~$D$
(see for instance
\cite{MM-lagsym}
\cite{CLM-convNoether}
and references therein)
if $\vend \comp [X,D] = 0$.
From any vector field $X$ one can construct a newtonoid vector field 
---with respect to~$D$--- as
$X + \vend \comp [D,X]$.
This construction,
which has been used in several papers to study the symmetries
of lagrangian dynamics, 
is a kind of generalisation of the complete lift
of a vector field on $Q$ to $\Tan Q$.
From equation (\ref{Y-reg}) it is then easy to deduce the following result: 

\begin{cor}
If the lagrangian is hyperregular then
$Y_h$ is a newtonoid vector field 
with respect to the dynamical vector field $\XL_o$
of velocity space,
and is the newtonoid vector field defined from the vector field 
$X_{\Leg^*(h)} = \Del_h$.
\qed
\end{cor}

In the singular case, 
using (\ref{XL-Leg}) it is readily seen that
$Y_h$ satisfies the condition of being newtonoid with respect to $\XL_o$
only on the primary lagrangian constraint submanifold~$V_1$.

\section{An example}

As a simple example, let us consider the lagrangian
of the conformal particle
\cite{Sie-spinpart}
\cite{GR-conformal}
\beq
L = {1\over2} (\dot x^2 - \lambda x^2) ,
\eeq
with configuration variables
$(x,\lambda) \in Q = \Real^n \times \Real$,
and $\Real^n$ endowed with an indefinite scalar product.
The Legendre transformation is given by
\beq
\Leg(x,\lambda;\dot x,\dot \lambda) = (x,\lambda;\hat p,\hat \pi) ,
\quad
\hat p = \dot x
,\;
\hat \pi = 0 ,
\eeq
so the primary constraint submanifold $P_o \subset \Tan^*Q$ 
has codimension~1, and is described by the primary hamiltonian constraint
\beq
\phi = \pi .
\eeq
As a hamiltonian we take
\beq
H = {1\over2} (p^2 + \lambda x^2) .
\eeq

Stabilization of $\phi^0 = \phi$ yields
three additional generations of constraints
$\phi^{i+1} = \{\phi^i,H\}$:
$$
\phi^1 = - \frac12 x^2 ,
\quad
\phi^2 = - p x ,
\quad
\phi^3 = \lambda x^2 - p^2 ,
$$
which are first-class.
The lagrangian constraints are
$\chi^i := K \lop \phi^{i-1}$:
$$
\chi = \chi^1 = - \frac12 x^2 ,
\quad
\chi^2 = - \dot x x ,
\quad
\chi^3 = \lambda x^2 - \dot x^2 .
$$
(Indeed $\chi^i = \Leg^*(\phi^i)$,
since the hamiltonian constraints are first-class.)
Notice also that
$
K \lop \phi^3 = -2 \dot\lambda \chi^1 - 4 \lambda \chi^2
$.

The kernel of $\Tan(\Leg)$ is spanned by 
$\Gam_\phi = \derpar{}{\dot \lambda}$.
From the identity $\Id = \gam_H + \lam \, \gam_\phi$
we determine the function $\lam = \dot\lambda$.
We also obtain
\beann
K \lop g
&=&
\dot x^a \,\Leg^*\!\left(\derpar g{x^a} \right) +
\dot \lambda \,\Leg^*\!\left(\derpar g\lambda \right) -
\lambda x_a \,\Leg^*\!\left(\derpar g{p_a} \right) -
\frac12 x^2 \,\Leg^*\!\left(\derpar g\pi \right)
\\
&=&
\Leg^*{\{g,H\}} + \Leg^*{\{g,\pi\}} \,\dot\lambda .
\eeann

Now we can compute
$\ds
Y_h = 
\Leg^*\!\!\left(\derpar{h}{p}\right) \derpar{}{x} +
\Leg^*\!\!\left(\derpar{h}{\pi}\right) \derpar{}{\lambda} +
\left(K \lop \derpar{h}{p}\right) \derpar{}{\dot x} +
\left(K \lop \derpar{h}{\pi}\right) \derpar{}{\dot \lambda} 
$,
and in particular 
$$
Y_\phi = \derpar{}{\lambda},
\qquad
Y_H = \dot x \derpar{}{x} - \lambda x \derpar{}{\dot x} .
$$

Then, from
$
R_h = \Gam_{\{h,H\}} + \dot\lambda \,\Gam_{\{h,\pi\}}
$
we get
$R_\phi = \Gam_{\phi^1} = 0$
and
$R_H = \dot \lambda \,\Gam_{-\phi^1} = 0$,
from which 
$\Del_\phi = Y_\phi$ and $\Del_H = Y_H$.

According to our results,
the kernel of the presymplectic form $\omega_L$ is spanned by
$\Gam_\phi = \derpar{}{\dot\lambda}$ and
$\Del_\phi = \derpar{}{\lambda}$.
(In this case this is obvious since $\omega_L = \dif x \wedge \dif \dot x$.)

Finally we get the primary dynamical vector fields as
$\XL = \XL_o + \eps \Gam_\phi$, where
$$
\XL_o = Y_H + \dot\lambda \,Y_\phi =
\dot x \derpar{}{x} + \dot\lambda \derpar{}{\lambda} 
- \lambda x \derpar{}{\dot x} .
$$
It is easily checked that
$\ds
\Tan(\Leg) \comp \XL_o - K = -\chi \, \derpar{}{\pi} \feble{} 0
$.

\section{Conclusions}

During the last two decades many papers have studied
the close relations between lagrangian and hamiltonian formalisms
when the lagrangian function is singular.
One can expedite the lagrangian picture by using
some results from the hamiltonian side.

In this paper we have added new objects to the geometric framework
of these relations.
First, for any function $h$ on phase space $\Tan^*Q$
we have defined the vector field $Y_h$ on velocity space $\Tan Q$.
When looked in coordinates,
this object reminds one of the definition of newtonoid vector fields;
but instead of using a second-order dynamics on~$Q$, 
which is not well defined in general when the lagrangian is singular,
we use the unambiguous time-evolution operator $K$ that
connects lagrangian and hamiltonian formalisms.
Once a hamiltonian $H$ and 
a set of primary hamiltonian constraints $\phi_\mu$
have been chosen,
we have also defined the vector fields $R_h$ and $\Del_h$.

These objects give effective answers to several questions.
The projectability of a vector field to a hamiltonian vector field:
we have shown that, when $h$ is a first-class function on $\Tan^*Q$,
the vector field $\Del_h$ projects to the hamiltonian vector field
$Z_h$.
The kernel of 
the presymplectic form of lagrangian formalism:
it can be computed as the subbundle spanned by the vector fields
$\Gam_\mu$ associated with the primary hamiltonian constraints $\phi_\mu$
and the vector fields 
$\Del_{\mu_o}$ associated with the first-class primary hamiltonian constraints.
The construction of the dynamical vector fields in lagrangian formalism:
the vector field
$\XL_o = \Del_H + \lam^\mu \Del_\mu$
is a solution of the Euler-Lagrange equation
on the primary lagrangian constraint submanifold.
Finally, the characterisation of dynamical symmetries:
the fact that $G$ is the generator of an infinitesimal symmetry
can be expressed as a kind of commutation relation between
the time-evolution operator~$K$
and the couple of vector fields
$Y_G$, $Z_G$.

In view of these results, we can say that the time-evolution operator~$K$
still provides one with new insights about the connections
between singular lagrangian and hamiltonian dynamics.
The functions $\lam^\mu$, given by (\ref{lam}) as a kind of
pseudo-inversion of the Legendre transformation,
and the fibre derivation,
a seldom used operation in geometric mechanics,
complete, 
together with the usual structures of tangent and cotangent bundles, 
the set of tools used in this paper.

As a final remark, 
let us point out that 
some of our expressions are also valid in the time-dependent case,
which is especially interesting for dealing with gauge symmetries.

\section*{Acknowledgments}
X.\,G. acknowledges financial support by CICYT projects 
TAP 97-0969-C03 and PB98--0920.
J.\,M.\,P. acknowledges financial support by 
CICYT, AEN98-0431, and CIRIT, GC 1998SGR.



\end{document}